\documentclass[10pt,twocolumn,letterpaper]{article}

\usepackage[pagenumbers]{wacv} 

\definecolor{wacvblue}{rgb}{0.21,0.49,0.74}
\usepackage[pagebackref,breaklinks,colorlinks,allcolors=wacvblue]{hyperref}

\def\wacvPaperID{20} 
\def\confName{WACV}
\def\confYear{2027}

\title{Scribble Once, Segment the Sequence: Bidirectional Temporal Consistency for Weakly Supervised Segmentation in Echocardiography Sequences}

\author{Vincent Li Weixin\\
University of Oxford\\
{\tt\small vincent.li@reuben.ox.ac.uk}
\and
Chen Jingkun\\
Northwest Polytechnical University Xi'an\\
{\tt\small jingkunchen@nwpu.edu.cn}
\and
Vicente Grau\\
University of Oxford\\
{\tt\small vicente.grau@eng.ox.ac.uk}
}

\begin{document}
\maketitle
\begin{abstract}
Temporal medical image segmentation relies on pixel-wise annotations, which are prohibitively expensive for long image sequences. Existing weakly-supervised methods reduce the annotation burden per frame but still require repeated annotations across the sequence. Furthermore, sparse supervision can result in temporally unstable representations. Addressing both challenges requires a model that can leverage strong spatial priors from large-scale pretraining while minimizing reliance on dense annotations. Frozen 2D vision foundation models provide a promising basis for this by offering transferable spatial representations with low trainable parameters. However, their lack of explicit temporal modeling limits their ability to propagate reliable information across anatomical sequences. To address this, we present a weakly-supervised framework for temporal segmentation from a single scribble-annotated frame, with all remaining frames left unlabeled. The framework combines parameter-efficient adaptation of DINOv3 encoders present in the mean-teacher architecture. Furthermore, a set of complementary losses, including a temporal loss helps generate reliable segmentation priors for the unlabeled frames. This formulation decouples annotation costs from sequence length while retaining the transferable spatial knowledge of the pretrained encoder. Extensive experiments on the publicly available CAMUS and CardiacNet datasets demonstrate that our approach outperforms other scribble-supervised methods. These results show that accurate and temporally consistent segmentation can be achieved from extremely sparse supervision, substantially reducing annotation requirements for temporal echocardiographic segmentation and providing a practical framework for scalable clinical deployment.
\end{abstract}
    
\section{Introduction}
\label{sec:introduction}

Segmentation of temporal medical sequences is vital in many clinical applications, as it enables consistent quantification and tracking of anatomical and pathological changes over time~\cite{chen2020deep, yang2025contrast, srikijkasemwat2025multi}. However, achieving reliable sequence segmentation remains constrained by the annotation bottleneck. Methods achieving high segmentation performance conventionally rely on dense pixel-wise annotations for every frame, while obtaining such large-scale annotations in clinical settings remains a significant challenge~\cite{wang2021annotation, tajbakhsh2020embracing}. This challenge is particularly pronounced in echocardiography, where speckle noise obscures anatomical boundaries and adjacent tissues can exhibit similar acoustic properties, making structures difficult to distinguish without expert interpretation~\cite{chen2026automatic, yang2026motion}. Consequently, annotation of echocardiographic sequences is both time-consuming and dependent on specialist expertise, with annotation effort increasing with every additional frame in a sequence.

Current methods that reduce annotation requirements can broadly be divided into two categories. Semi-supervised learning (SSL) combines a small subset of densely annotated frames with a larger set of unlabeled frames, as shown in Mean Teacher~\cite{tarvainen2017mean} and Cross Pseudo Supervision (CPS)~\cite{chen2021semi}. The labeled subset provides direct supervision through segmentation losses, while the unlabeled subset provides consistency regularization~\cite{lu2023mutually} that encourages reliable and temporally coherent predictions. Weakly-supervised learning (WSL), in contrast, replaces dense masks with coarse annotations such as bounding boxes~\cite{lan2021discobox, rajchl2016deepcut, kervadec2020bounding}, points~\cite{zhao2020weakly, zhai2023pa} or scribbles~\cite{chen2022scribble2d5, can2018learning, chen2025volumetric}, typically provided on every training frame. These approaches reduce annotation burden in different ways. SSL reduces the number of frames requiring dense annotation while WSL reduces the amount of annotation required per frame. However, both remain dependent on annotations across the sequence, so annotation effort continues to increase with sequence length.

To address this limitation, we propose a weakly-supervised framework for segmenting echocardiographic sequences from a single scribble-annotated frame, with all remaining frames left unlabeled. The framework consists of three complementary components that convert sparse spatial supervision into reliable sequence-wide segmentation. Firstly, a mean-teacher framework generates dense pseudo-labels and propagates supervision outward from the annotated frame in a bidirectional manner. Secondly, a frozen DINOv3-based~\cite{simeoni2025dinov3} encoder, adapted to medical images using Low-rank Adaptations (LoRA)~\cite{hu2021lora} provides transferable spatial representations while restricting adaptation to a small number of trainable parameters. The representations are fed into a lightweight U-Net-style decoder~\cite{ronneberger2015u}, keeping trainable parameters low. Thirdly, complementary spatial and temporal losses enforce consistency across the sequence while allowing anatomically plausible changes in cardiac shape.

We evaluate the proposed framework on two publicly available datasets, Cardiac Acquisitions for Multi-structure Ultrasound Segmentation (CAMUS)~\cite{leclerc2019deep} and CardiacNet~\cite{yang2024cardiacnet}, spanning three different cohorts. Experimental results demonstrate that our method consistently outperforms existing scribble-supervised baselines in both Dice and HD95 metrics, while achieving a more favorable balance between precision and recall. These results show that accurate and spatially well-localized temporal segmentation can be achieved from only one scribble-annotated frame, substantially reducing annotation requirements for echocardiographic sequence segmentation. Our key contributions are summarized as follows:

\begin{itemize}
    \item We propose a weakly-supervised temporal segmentation framework for medical imaging sequences based on single slice scribble annotations on the middle frame.
    \item We demonstrate that a LoRA-adapted foundation model encoder can improve segmentation performance with significantly fewer parameters compared to state-of-the-art scribble segmentation methods
    \item We also propose an anchor-relative bidirectional temporal loss, $L_{temp}$, that can propagate supervision outward from the anchor frame in both directions while allowing for drifting so that segmentation masks follow the natural progression of anatomical structures.
\end{itemize}
\section{Related Work}
\label{sec:related_work}

\subsection{Scribble-supervised Segmentation}

Scribble-supervised segmentation addresses the problem of learning dense segmentation masks from sparse pixel annotations. Common methods such as partial cross-entropy (pCE) compute the supervised loss only over annotated pixels but do not extend supervision to unlabeled regions. Therefore, existing scribble-supervised methods extend sparse scribbles through augmentation, consistency regularization and architectural modifications.

Augmentation-based methods perturb the training samples and enforce consistency between the corresponding predictions. CutMix~\cite{yun2019cutmix} constructs a new training sample by combining regions from two images, with the corresponding labels mixed using the same spatial transformation. This exposes the model to additional combinations of foreground and background regions while preserving the available sparse supervision. CycleMix~\cite{zhang2022cyclemix} extends mixed-sample augmentation with cycle-consistency constraints, using complementary transformed views to exploit information from both labeled and unlabeled regions. Together, these strategies encourage predictions to remain consistent under input perturbations.

Architecture-based approaches instead modify the segmentation framework to better exploit the limited spatial supervision. ScribFormer~\cite{li2024scribformer} combines a convolutional branch with a transformer branch and a class-activation branch to integrate local appearance with global contextual information, allowing the network to infer object structure beyond the annotated scribbles. Other methods combine architectural changes with consistency-based training. HELPNet~\cite{zhang2025helpnet} introduces hierarchical perturbations at multiple spatial granularity and enforces consistency between the resulting predictions, together with edge-guided refinement to improve boundaries that are not directly specified by the scribble. Despite different mechanisms, these methods process frames independently and require scribble annotations throughout the sequence, causing annotation cost to scale with sequence length.

\subsection{Vision Foundation Models and Parameter-Efficient Adaptation}

Large-scale self-supervised pretraining has produced vision backbones with transferable representations that can be reused across downstream tasks with limited task-specific supervision~\cite{li2023systematic}. DINOv3 extends this capability through large-scale self-supervised pretraining, providing spatial representations that can be transferred to domains where labeled data are scarce. We use a general-purpose DINOv3 rather than an echocardiography-specific foundation model so that the supervision scheme remains modality-agnostic rather than tied to a single domain.

Adapting large pretrained backbones to a target medical domain introduces a second challenge. Full fine-tuning of all backbone parameters increases computational cost and risk of overfitting~\cite{silva2025towards}. Parameter-efficient fine-tuning (PEFT) methods instead preserve most of the pretrained representation while restricting task-specific adaptation to a small number of parameters. LoRA introduces trainable low-rank updates into selected linear projections while keeping the original pretrained weights fixed, substantially reducing the number of trainable parameters.
\section{Methodology}
\label{sec:methodology}

\subsection{Overview}

\begin{figure*}[t]
  \centering
  \includegraphics[width=\linewidth]{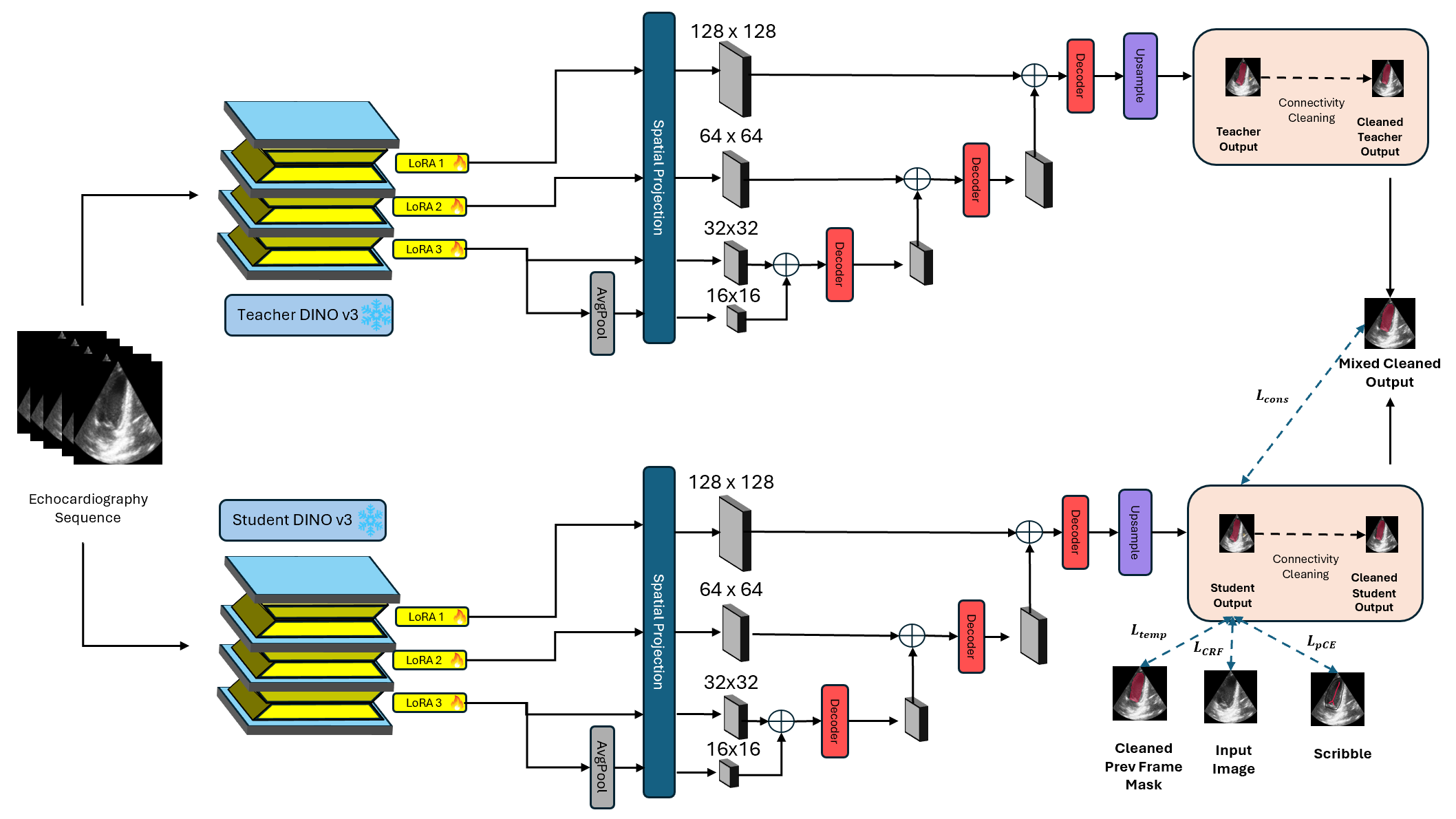}
  \caption{Overview of the proposed weakly-supervised temporal segmentation framework. A mean-teacher architecture with LoRA-adapted frozen DINOv3 encoders generates mixed connectivity-filtered pseudo-labels for dense Dice supervision. Partial cross-entropy supervises scribbles on the anchor frame, gated-CRF refines boundaries and an anchor-relative temporal loss enforces bidirectional consistency while allowing cardiac deformation}
  \label{fig:architecture}
\end{figure*}

As illustrated in~\cref{fig:architecture}, we propose a weakly-supervised framework for segmenting echocardiographic sequences from scribble annotations on the anchor frame. The following sections describe each component in detail.

\subsection{Segmentation of Echocardiographic Sequences from a Single Scribble}

Let $X = \{x_1, x_2, \ldots, x_T\}$ denote an echocardiographic sequence
of $T$ frames. The model predicts a segmentation mask for every frame with supervision limited to a single scribble map $S$ defined on the anchor frame $x_m$ while all remaining frames are unlabeled. The annotation map contains both foreground and background scribbles and the scribble assigns each pixel $i$ of $x_m$ one of three values

\begin{equation}
  S(i) =
  \begin{cases}
    0,   & \text{background},\\
    1,   & \text{foreground},\\
    255, & \text{ignore}
  \end{cases}
  \label{eq:scribble}
\end{equation}

The framework allows any frame in the sequence to be selected as the annotated anchor frame. The middle frame $x_m$ provides a natural default by reducing the maximum propagation distance to approximately $T/2$ frames. However, keyframes such as end-systole (ES) or end-diastole (ED) frames may also be selected. Supervision is subsequently propagated bidirectionally from the selected anchor to the remaining unlabeled frames.

\subsection{Parameter-Efficient Encoding and Multi-Scale Decoding}

A frozen DINOv3 encoder is adapted using LoRA on the query and value projections of three transformer blocks, introducing $27.6$\,K trainable parameters corresponding to $0.03\%$ of the $85.7$\,M while preserving the pretrained representation under sparse supervision. Patch tokens are extracted from the three adapted blocks. Each token corresponds to a $16\times16$ pixel patch so a $512\times512$ input yields a grid of $32\times32$ tokens. Since the transformer block applies no spatial downsampling, all three LoRA block readouts share this resolution and differ only in semantic depth. They are resampled to strides $4$, $8$ and $16$, producing $128\times128$, $64\times64$ and $32\times32$ spatial feature maps. A fourth level at stride $32$ is obtained by average pooling the stride-$16$ map to $16\times16$, which provides the decoder a coarser representation and a larger effective receptive field.

Each pyramid level is independently projected using a $1\times1$ convolution followed by group normalization and ReLU. The projected features are then passed to a U-Net style decoder, beginning at stride-$32$ level and progressively upsampled through three stages. At each stage, the upsampled decoder features are concatenated with the corresponding projected feature map from the pyramid. The final decoder output is converted to segmentation logits and bilinearly resized to the original input resolution.

\subsection{Dense Supervision from Single Scribble-annotated Frame}

\subsubsection{Mean Teacher Construction and Scribble Supervision}
The framework maintains structurally identical student and teacher branches. The student is optimized directly through gradient descent while the teacher parameters are updated as an exponential moving average (EMA) of the student parameters. Since the DINOv3 backbone is frozen throughout training, its pretrained weights are shared between the two branches, while the LoRA adapters, feature projections and segmentation decoder are maintained separately and updated through the corresponding student and teacher mechanisms.

For each training sequence, the student receives a weakly augmented view generated using brightness and contrast perturbations together with additive noise. These transformations modify image appearance while preserving spatial geometry, ensuring pixel-wise alignment between the student predictions and the supervision generated from the unaugmented sequence.

Direct supervision is restricted to the annotated frame $x_m$, which is the only frame containing scribble annotations. We apply pCE exclusively to pixels labeled as foreground and background

\begin{equation}
  \mathcal{L}_{\mathrm{pCE}}
  = -\frac{1}{|\Omega|}\sum_{i\in\Omega}\sum_{c=0}^{C-1}
    y_i^{c}\,\log p_i^{c}
  \label{eq:pce}
\end{equation}

\noindent where $\Omega = \{\, i : S(i) \neq 255 \,\}$ is the set of annotated pixels on the middle frame, $y_i^{c}$ is the one-hot encoding of the scribble label at pixel $i$ and $p_i^{c}$ is the predicted probability of class $c$ at that pixel. Pixels
assigned the ignore value $255$ lie outside $\Omega$ and are excluded from the loss. This ensures that supervision is applied only where an explicit annotation is available, without assigning artificial labels to the remaining pixels in the scribble-annotated frame.

\subsubsection{Dense Pseudo-label Construction}

We use spatial connectivity to the foreground scribble to identify the predicted component associated with the annotated structure, and propagate this selection through the sequence. Let $\mathrm{CC}(\hat{y})$ denote the set of connected components of a predicted binary mask $\hat{y}$, so that each $\mathcal{R} \in \mathrm{CC}(\hat{y})$ is a single maximally connected foreground region, and let $\mathcal{S}_{\mathrm{fg}} = \{\, i : S(i) = 1 \,\}$ denote the set of foreground scribble pixels. The connectivity filter $\Phi(\cdot)$ maps a predicted mask to the subset that remains associated with the annotation. For the anchor-slice prediction $\hat{y}_m$, only components intersecting the foreground scribble are retained

\begin{equation}
  \Phi(\hat{y}_m)
  = \bigcup \bigl\{\, \mathcal{R} \in \mathrm{CC}(\hat{y}_m)
    \;:\; \mathcal{R} \cap \mathcal{S}_{\mathrm{fg}} \neq \emptyset \,\bigr\}
  \label{eq:cc-anchor}
\end{equation}

Starting from the anchor slice, the filtered mask is propagated outward in both temporal directions. Each frame retains only those connected components that overlap the already filtered prediction of its adjacent frame, $\hat{y}_t$

\begin{equation}
  \Phi(\hat{y}_t)
  = \bigcup \bigl\{\, \mathcal{R} \in \mathrm{CC}(\hat{y}_t)
    \;:\; \mathcal{R} \cap \Phi(\hat{y}_{t \mp 1}) \neq \emptyset \,\bigr\}
  \label{eq:cc-prop}
\end{equation}

Connectivity filtering is applied independently to both the teacher and student predictions before fusion, ensuring that each branch retains only components connected to the scribble-propagated structure. Let $\Phi^{T}_t \equiv \Phi(\hat{y}^{T}_t)$ and $\Phi^{S}_t \equiv \Phi(\hat{y}^{S}_t)$ denote the connectivity-filtered teacher and student predictions at frame $t$ respectively. These predictions are then combined through pixel-wise mixing~\cite{luo2022scribble}, incorporating the more stable EMA teacher prediction with the faster-adapting student prediction. Let $M_t \in \{0,1\}^{H \times W}$ be a random selection mask that assigns each pixel to one of the two branches, drawn independently from $\mathrm{Bernoulli}(\beta)$. The mixed pseudo-label is 

\begin{equation}
  \tilde{y}_t = M_t \odot \Phi^{T}_t + (\mathbf{1} - M_t) \odot \Phi^{S}_t
  \label{eq:mix}
\end{equation}

\noindent where $\odot$ denotes the element-wise product, so each pixel of $\tilde{y}_t$ is taken from either branch. The student prediction is supervised against $\tilde{y}$ on every frame by a soft Dice loss $L_{cons}$ averaged over both classes.

\subsubsection{Anchor-relative Temporal Consistency}

A temporal consistency loss, $\mathcal{L}_{\text{temp}}$, encourages coherent predictions across neighboring frames while preserving outward propagation from the anchor frame $x_m$. We define a temporal window $\mathcal{D} = \{\, 1, 2, \ldots, D \,\}$, where $d \in \mathcal{D}$ denotes the frame distance. For each frame $t$, the references are the $D$ frames adjacent to $t$ on the side facing the anchor. Frames before the anchor take references after them and frames at or after the anchor take references before them

\begin{equation}
  r = \begin{cases} t + d, & t < m, \\ t - d, & t \ge m, \end{cases}
  \qquad d \in \mathcal{D}.
  \label{eq:temp_ref}
\end{equation}

A frame far from the anchor is compared with its own neighbors rather than with the anchor. Each reference is weighted by $w_d = 1/d$, giving greater influence to nearby frames. This progressively propagates supervision without enforcing the anchor geometry across the entire sequence. To avoid suppressing natural anatomical changes between frames, three mechanisms are introduced to distinguish physiological variation from inconsistent predictions.

Firstly, a confidence gate restricts the temporal loss to pixels where the student is confident and agrees with the pseudo-label of the reference frame. Let $\hat{y}_t(i)$ be the class the student predicts at pixel $i$ of frame $t$, and $\hat{p}_t(i)$ the probability it assigns to that class. The gate $\Omega_{t,r}$ is the set of pixels of frame $t$ that are kept when it is compared with its reference frame $r$

\begin{equation}
  \Omega_{t,r} = \{\, i \;:\; \hat{p}_t(i) > \theta \ \text{and}\ \hat{y}_t(i) = \tilde{y}_r(i) \,\}.
  \label{eq:temp_gate}
\end{equation}

where $\theta$ is a confidence threshold where a pixel is kept only if the student assigns its predicted class with probability above $\theta$. Pixels outside $\Omega_{t,r}$ are ignored and a frame pair contributes to the temporal loss only when $\Omega_{t,r}$ covers at least a fraction $\eta$ of the frame. This prevents uncertain or conflicting predictions from enforcing temporal consistency. Secondly, predictions are average-pooled over non-overlapping cells before comparison rather than evaluated at the original resolution. Each pooled value therefore represents the foreground occupancy within a local region allowing small spatial displacements between neighboring frames without penalty. Thirdly, a margin $\gamma$ is applied to the difference between the pooled prediction and pseudo-label

\begin{equation}
  e_{t,r}(u) = \max\!\left( \left| \bar{p}_t(u) - \bar{y}_r(u) \right| - \gamma,\; 0 \right),
  \label{eq:temp_deadzone}
\end{equation}

where $e_{t,r}(u)$ is the error assigned to cell $u$ when frame $t$ is compared with its reference $r$ and $\left| \bar{p}_t(u) - \bar{y}_r(u) \right|$ measures how much the foreground occupancy of cell $u$ differs between the two frames. Differences smaller than $\gamma$ incur no penalty while larger differences are penalized only by the amount that exceeds $\gamma$.

\subsubsection{Boundary Regularization and Training Objective}

A gated conditional random field (CRF) loss, $L_{CRF}$, grounds the predicted boundaries in the underlying image appearance. Pixels that are spatially close and similar in intensity are encouraged to receive the same label, while disagreement across an intensity edge is left unpenalized. This supplies complementary supervision in the unlabeled region between the foreground and background strokes. The complete objective combines the four terms 

\begin{equation}
  \mathcal{L} =\lambda_{\mathrm{pCE}} \mathcal{L}_{\mathrm{pCE}}
  + \lambda_{\mathrm{cons}}\mathcal{L}_{\mathrm{cons}}
  + \lambda_{\mathrm{temp}}\mathcal{L}_{\mathrm{temp}}
  + \lambda_{\mathrm{CRF}}\mathcal{L}_{\mathrm{CRF}}
  \label{eq:total}
\end{equation}

\noindent where $\mathcal{L}_{\mathrm{pCE}}$ applies only at the annotated frame $x_m$, while the remaining three apply across the whole sequence. 
\section{Experiments and Results}
\label{sec:experiments_and_results}

\subsection{Datasets}

We evaluate our method on two publicly available echocardiography datasets, CAMUS and CardiacNet. All datasets are divided into training, validation and test sets using a 70:10:20 ratio. CAMUS contains echocardiographic sequences from 500 patients, from which we use the two-chamber (2CH) half-sequence subset. Ground-truth masks are provided for all frames, so the mask-derived scribble annotations is on the middle frame of the sequence. Performance is evaluated on three anatomical structures namely the left-ventricular endocardium ($LV_{endo}$), left-ventricular epicardium ($LV_{epi}$) and left atrium.

CardiacNet comprises two pathology-specific cohorts: atrial septal defect (ASD) and pulmonary arterial hypertension (PAH). Segmentation masks are released for only a subset of the acquired sequences and a small number of the released files could not be read. We therefore use every sequence for which a usable annotation exists, giving 135 ASD and 299 PAH sequences. Unlike CAMUS, ground-truth masks are provided for approximately five frames within each sequence, so the scribble is placed on the middle annotated frame rather than the middle frame of the sequence. Both cohorts are evaluated on four cardiac structures: left ventricle (LV), left atrium (LA), right ventricle (RV) and right atrium (RA). Across both datasets, training supervision is restricted to one background and one foreground scribble with all remaining frames unlabeled.

\subsection{Implementation Details}

All experiments were implemented in PyTorch with Python 3.12 and CUDA 13.2 and conducted on a single NVIDIA RTX PRO 5000 Blackwell GPU. We use DINOv3 ViT-B/16 as the vision foundation model backbone, with rank-3 LoRA adapters applied to the query and value projections of transformer blocks 3, 6 and 9. Single-channel ultrasound frames are used across three channels, resized to $512\times512$, and normalized using the ImageNet channel-wise mean and standard deviation. 

Models are trained for 40 epochs using AdamW with a learning rate of $1\times10^{-4}$, weight decay of $1\times10^{-4}$ and mixed-precision training. The partial cross-entropy, Dice, temporal consistency and gated-CRF losses are weighted by $1.0$, $0.5$, $0.1$ and $0.1$ respectively. The weights of the three unsupervised objectives are linearly ramped from zero to their respective values over the first five epochs. The teacher parameters are updated using an EMA decay of $0.99$ and the connectivity-filtered teacher and student predictions are mixed using a pixel-wise Bernoulli mask with $\beta=0.5$.

\subsection{Baseline and Evaluation Metrics}

We compare our method against a U-Net trained with pCE alone and four scribble-supervised methods: CutMix, CycleMix, ScribFormer and HELPNet. A consistent evaluation protocol is applied across all methods, with performance measured on all frames with available ground-truth annotations to assess generalization beyond the annotated frame. We report the Dice similarity coefficient and 95th-percentile Hausdorff distance (HD95), together with precision and recall to further characterize segmentation errors.

\begin{table*}[t]
  \centering
  \footnotesize
  \setlength{\tabcolsep}{3.2pt}
  \caption{Segmentation results for all eleven structures. Top: Dice similarity
  coefficient ($\uparrow$). Bottom: 95th-percentile Hausdorff distance in pixels
  ($\downarrow$). Avg is the mean over the structures of each cohort.
  \#Params: trainable parameters in millions.}
  \label{tab:results}
  \begin{tabular}{l c cccc ccccc ccccc}
    \toprule
    & & \multicolumn{4}{c}{CAMUS} & \multicolumn{5}{c}{CardiacNet-ASD}
    & \multicolumn{5}{c}{CardiacNet-PAH}\\
    \cmidrule(lr){3-6}\cmidrule(lr){7-11}\cmidrule(lr){12-16}
    Method & Params (M) & $LV_{endo}$ & $LV_{epi}$ & LA & Avg
           & LV & LA & RV & RA & Avg
           & LV & LA & RV & RA & Avg\\
    \midrule
    \multicolumn{16}{l}{\textit{Dice ($\uparrow$)}}\\
    \midrule
    U-Net (pCE)   & 7.76           & 0.351 & 0.635 & 0.337 & 0.441
                                   & 0.057 & 0.049 & 0.084 & 0.086 & 0.069
                                   & 0.198 & 0.190 & 0.297 & 0.127 & 0.203\\
    CutMix        & 7.76           & 0.650 & 0.709 & 0.702 & 0.687
                                   & 0.275 & 0.492 & 0.154 & 0.704 & 0.406
                                   & 0.390 & 0.647 & 0.455 & \textbf{0.757} & 0.562\\
    CycleMix      & 7.76           & 0.560 & 0.700 & 0.334 & 0.531
                                   & 0.177 & 0.054 & 0.093 & 0.072 & 0.099
                                   & 0.063 & 0.084 & 0.268 & 0.092 & 0.127\\
    ScribFormer   & 47.91          & 0.239 & 0.787 & 0.148 & 0.391
                                   & 0.074 & 0.066 & 0.097 & 0.071 & 0.077
                                   & 0.078 & 0.064 & 0.101 & 0.096 & 0.085\\
    HELPNet       & 25.76          & 0.653 & 0.720 & 0.571 & 0.648
                                   & 0.131 & 0.121 & 0.116 & 0.045 & 0.103
                                   & 0.058 & 0.060 & 0.122 & 0.044 & 0.071\\
    \textbf{Ours} & \textbf{3.81}  & \textbf{0.851} & \textbf{0.791} & \textbf{0.709} & \textbf{0.784}
                                   & \textbf{0.741} & \textbf{0.705} & \textbf{0.762} & \textbf{0.770} & \textbf{0.744}
                                   & \textbf{0.781} & \textbf{0.740} & \textbf{0.752} & 0.744 & \textbf{0.754}\\
    \midrule
    \multicolumn{16}{l}{\textit{HD95 in pixels ($\downarrow$)}}\\
    \midrule
    U-Net (pCE)   & 7.76           & 224.27 & 175.23 & 287.61 & 229.03
                                   & 306.74 & 337.79 & 305.20 & 314.50 & 316.06
                                   & 304.27 & 337.45 & 239.80 & 327.60 & 302.28\\
    CutMix        & 7.76           & 183.23 & 86.79 & 68.23 & 112.75
                                   & 268.65 & 362.15 & 279.17 & 106.27 & 254.06
                                   & 210.44 & 116.34 & 217.34 & 79.73 & 155.96\\
    CycleMix      & 7.76           & 193.54 & 131.95 & 287.99 & 204.49
                                   & 311.06 & 342.17 & 298.59 & 334.88 & 321.67
                                   & 305.32 & 322.31 & 250.00 & 324.55 & 300.54\\
    ScribFormer   & 47.91          & 253.77 & 54.67 & 392.91 & 233.78
                                   & 308.20 & 345.08 & 304.32 & 332.61 & 322.56
                                   & 301.42 & 350.29 & 304.77 & 338.55 & 323.76\\
    HELPNet       & 25.76          & 109.49 & \textbf{21.57} & 135.27 & 88.78
                                   & 224.32 & 230.10 & 272.46 & 332.83 & 264.93
                                   & 261.79 & 322.53 & 212.49 & 350.55 & 286.84\\
    \textbf{Ours} & \textbf{3.81}  & \textbf{29.01} & 24.31 & \textbf{40.44} & \textbf{31.25}
                                   & \textbf{43.01} & \textbf{38.31} & \textbf{42.66} & \textbf{42.89} & \textbf{41.72}
                                   & \textbf{30.87} & \textbf{24.56} & \textbf{34.19} & \textbf{23.59} & \textbf{28.30}\\
    \bottomrule
  \end{tabular}
\end{table*}

\cref{tab:results} shows that our method achieves the highest average Dice across all three cohorts, with scores of $0.784$, $0.744$, and $0.754$ on CAMUS, CardiacNet-ASD and CardiacNet-PAH respectively. Performance is lowest on the CardiacNet-ASD cohort, where the septal defect disrupts the normal boundary between the atria, increasing ambiguity in chamber delineation. Compared to consistency-based methods, our method improves over CutMix by $0.097$, $0.338$, and $0.192$ Dice respectively. The improvement over dedicated scribble-supervised methods is larger, particularly on CardiacNet. While ScribFormer and HELPNet achieve competitive performance on some CAMUS structures, their average Dice remains low on both CardiacNet cohorts, indicating poor generalization while our method achieves $0.744$ on CardiacNet-ASD and $0.754$ on CardiacNet-PAH. Overall, our method achieves the highest Dice on $10$ of $11$ structures, demonstrating more consistent performance under single-frame scribble supervision. It also shows that our method achieves the lowest average HD95 across all three cohorts, with $31.25$, $41.72$, and $28.30$ pixels on CAMUS, CardiacNet-ASD, and CardiacNet-PAH respectively. Consistent with the Dice results, CardiacNet-ASD remains the most challenging cohort, although our method maintains considerably lower boundary errors across all four structures. 

\begin{figure*}[t]
  \centering
  \includegraphics[width=\linewidth]{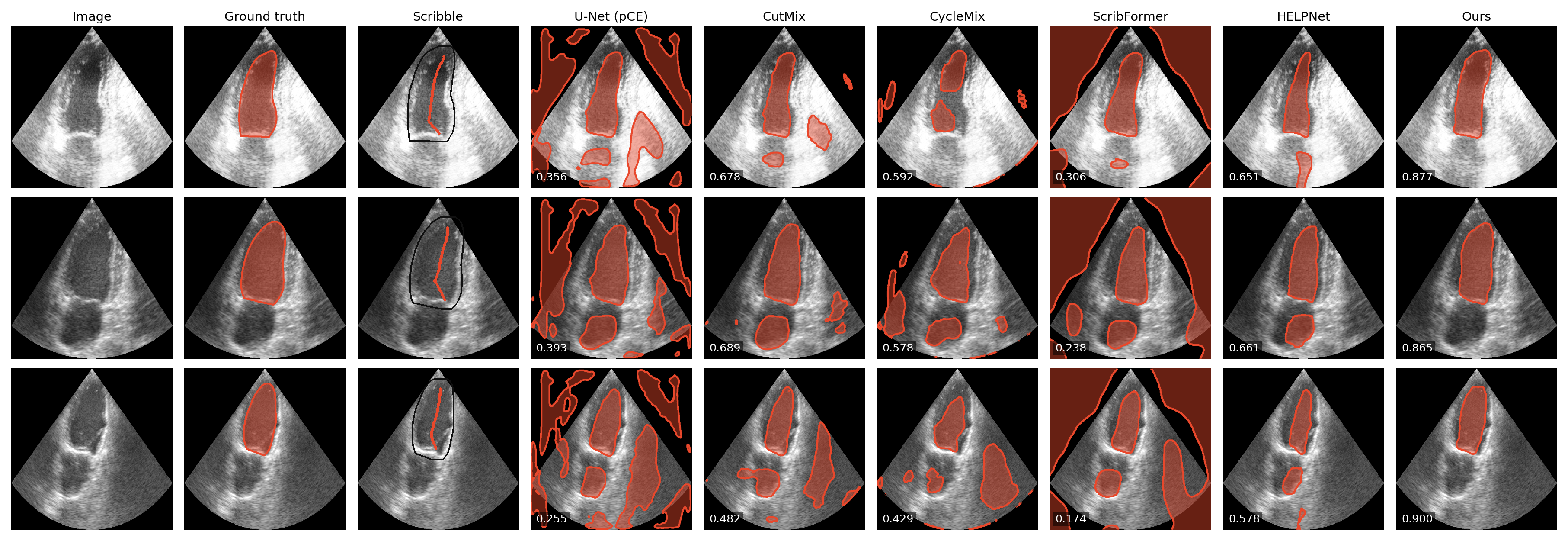}
  \caption{Qualitative comparison on CAMUS (LV endocardium) with the red segmentation mask overlayed on the image}
  \label{fig:qual-camus}
\end{figure*}

\begin{figure*}[t]
  \centering
  \includegraphics[width=\linewidth]{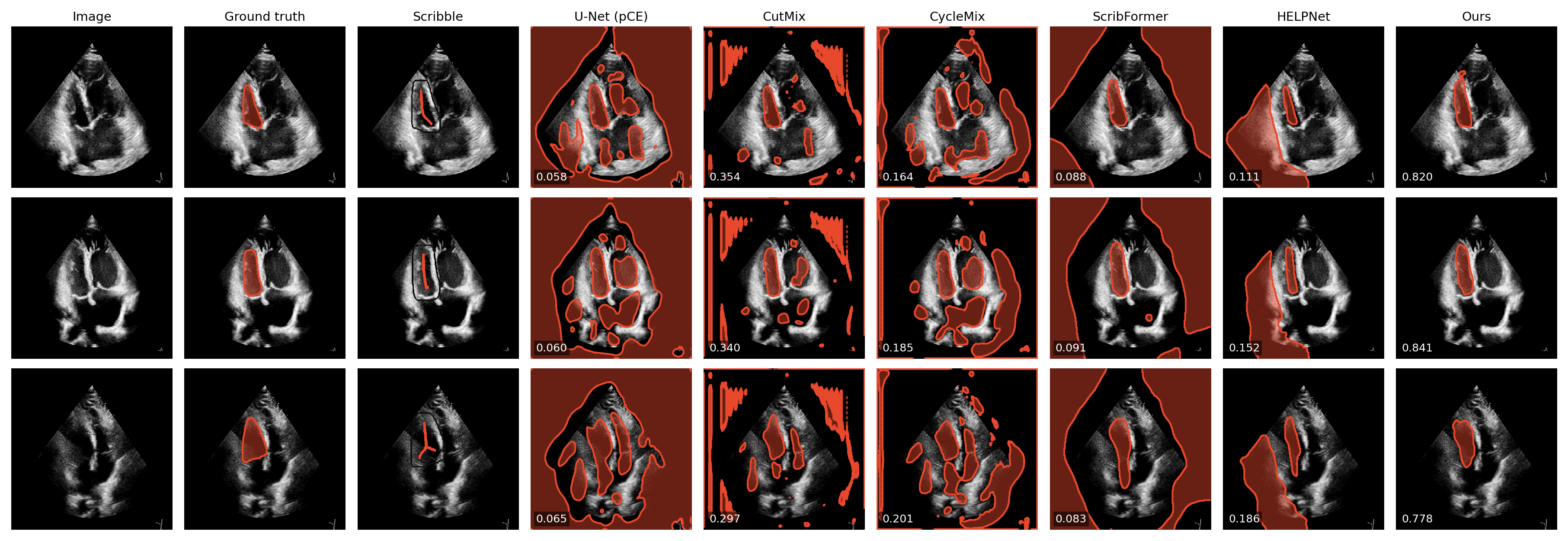}
  \caption{Qualitative comparison on CardiacNet-ASD (LV) with the red segmentation mask overlayed on the image}
  \label{fig:qual-asd}
\end{figure*}

\begin{figure*}[t]
  \centering
  \includegraphics[width=\linewidth]{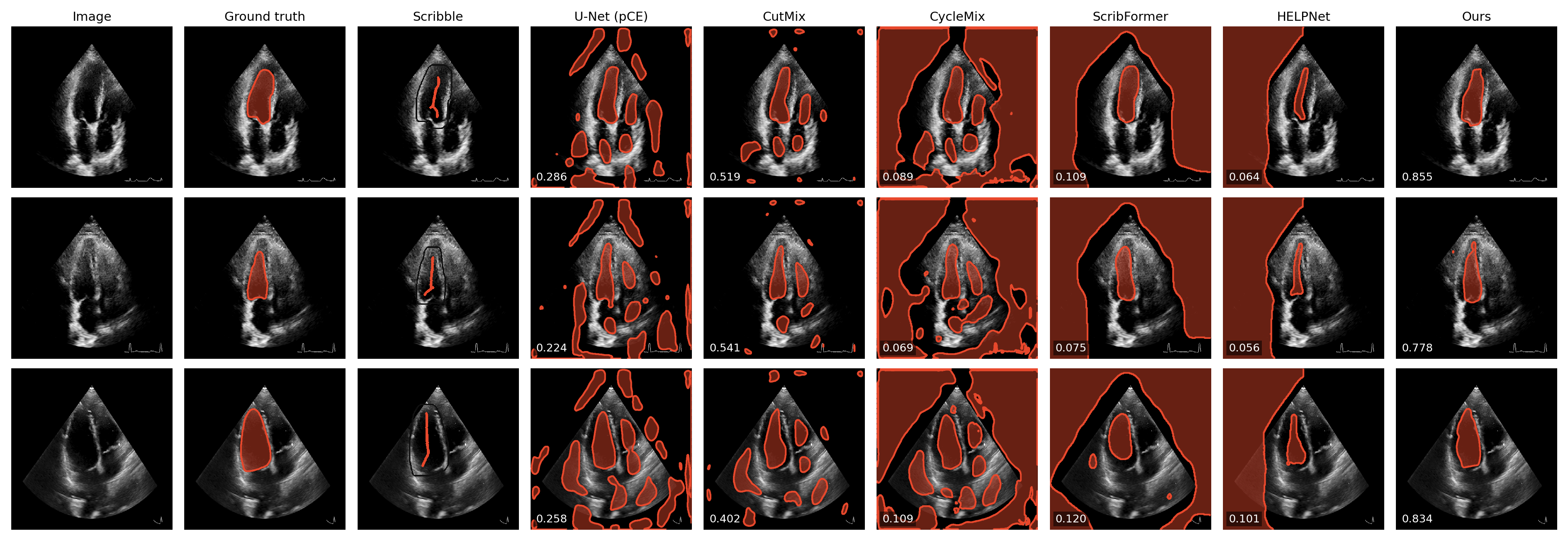}
  \caption{Qualitative comparison on CardiacNet-PAH (LV) with the red segmentation mask overlayed on the image}
  \label{fig:qual-pah}
\end{figure*}

Given the substantial performance gap observed in both Dice and HD95, we further investigate why existing methods degrade under single-frame supervision. As illustrated in~\cref{fig:qual-camus,fig:qual-asd,fig:qual-pah}, their poor performance is primarily driven by severe over-segmentation rather than localized boundary errors, with predictions extending substantially beyond the target anatomy. This explains the low Dice and high HD95 observed in the quantitative results. To better characterize this failure mode, we additionally analyze precision and recall in~\cref{tab:precrec}.

\begin{table}[t]
  \centering
  \footnotesize
  \setlength{\tabcolsep}{3pt}
  \caption{Precision / recall ($\uparrow$), averaged over the structures of each
  cohort. ASD and PAH are the two CardiacNet cohorts.}
  \label{tab:precrec}
  \begin{tabular}{lccc}
    \toprule
    Method & CAMUS & ASD & PAH\\
    \midrule
    U-Net (pCE) & 0.336 / 0.741 & 0.037 / 0.781 & 0.121 / 0.803\\
    CutMix      & 0.711 / 0.706 & 0.365 / 0.696 & 0.517 / 0.713\\
    CycleMix    & 0.474 / 0.699 & 0.055 / 0.748 & 0.074 / 0.770\\
    ScribFormer & 0.342 / 0.747 & 0.041 / 0.683 & 0.045 / 0.780\\
    HELPNet     & 0.746 / 0.605 & 0.062 / 0.390 & 0.041 / 0.350\\
    \midrule
    \textbf{Ours} & 0.901 / 0.713 & 0.780 / 0.743 & 0.858 / 0.698\\
    \bottomrule
  \end{tabular}
\end{table}

\cref{tab:precrec} further corroborates this failure mode. The pCE baseline, CycleMix and ScribFormer consistently exhibit high recall but extremely low precision, particularly on the two CardiacNet cohorts. This pattern shows over-segmentation, where predicted masks cover large regions of the image, leading to substantial overlap with the ground truth mask. Thus, recall values are inflated despite poor localization quality. In contrast, our framework improves precision while maintaining competitive recall, with precision consistently exceeding recall across all three cohorts. Overall, the improvements in Dice and HD95 can be attributed primarily to a reduction in false-positive predictions rather than increased anatomical coverage, consistent with the qualitative segmentation results.
\section{Ablations}

All ablation experiments are conducted on the CAMUS dataset using the LV endocardium. CAMUS is selected as ground-truth masks are available for every frame so metrics can be computed on the entire sequence, enabling a more comprehensive evaluation of temporal segmentation performance. The LV endocardium is used as a representative structure to systematically isolate the contribution of each component in the proposed framework. We evaluate the effects of the architectural design, LoRA adaptation, mean-teacher framework and individual loss terms.

\begin{table}[t]
  \centering
  \footnotesize
  \setlength{\tabcolsep}{4pt}
  \caption{Ablation study on CAMUS $LV_{\mathrm{endo}}$ and CardiacNet-PAH LV, reporting Dice ($\uparrow$). Each row differs from the full model in one component while loss rows are cumulative. Trainable parameters exclude the frozen $85.7$\,M backbone.}
  \label{tab:ablation}
  \begin{tabular}{lccc}
    \toprule
    Variant & Trainable & CAMUS & PAH\\
    \midrule
    \multicolumn{4}{l}{\textbf{Architecture}}\\
    \quad Three-level pyramid         & 3.00\,M  & 0.734 & 0.763\\
    \midrule
    \multicolumn{4}{l}{\textbf{LoRA}}\\
    \quad No LoRA                     & 3.78\,M  & 0.784 & 0.771\\
    \midrule
    \multicolumn{4}{l}{\textbf{Framework}}\\
    \quad Student only, no EMA        & 3.81\,M  & 0.799 & 0.772\\
    \midrule
    \multicolumn{4}{l}{\textbf{Losses}}\\
    \quad $\mathcal{L}_{\mathrm{pCE}}$ only & 3.81\,M & 0.613 & 0.410\\
    \quad $+\ \mathcal{L}_{\mathrm{cons}}$   & 3.81\,M & 0.770 & 0.626\\
    \quad $+\ \mathcal{L}_{\mathrm{temp}}$  & 3.81\,M & 0.812 & \textbf{0.790}\\
    \quad \textbf{$+\ \mathcal{L}_{\mathrm{CRF}}$ (Full Model)}
                                            & \textbf{3.81\,M} & \textbf{0.851} & 0.781\\
    \bottomrule
  \end{tabular}
\end{table}

\subsection{Effectiveness of Architecture}

We further examine the depth of the feature pyramid by reducing it from four to three levels, removing the coarsest scale while keeping the frozen DINOv3 encoder and all other components unchanged. The three-level model decreases Dice from $0.824$ to $0.734$ on CAMUS and from $0.779$ to $0.763$ on CardiacNet-PAH. The coarsest level aggregates information over a larger area of the image, allowing the decoder to reason about global shape and spatial relationships between structures, which cannot be inferred from local appearance alone. Without this level, the decoder must recover such context from finer scales that look at smaller areas, so spatially separated regions are processed largely independently. This is particularly detrimental under scribble supervision, where most pixels and unlabeled and the network must rely on global shape priors to extend sparse annotations to complete, anatomically plausible masks.

\subsection{Effectiveness of LoRA}

Removing the low-rank adapters reduces Dice from $0.824$ to $0.784$ on CAMUS and from $0.779$ to $0.771$ on CardiacNet-PAH. The improvement on CAMUS indicates that adapting the pretrained DINOv3 features is important for capturing echocardiographic characteristics such as low-contrast boundaries and speckle patterns. 

\subsection{Effectiveness of Mean Teacher Framework}

Removing the EMA teacher while retaining all other components reduces Dice from $0.824$ to $0.799$ on CAMUS and from $0.779$ to $0.772$ on CardiacNet-PAH. The consistent reduction across both datasets demonstrates the benefit of maintaining a temporally averaged teacher. Without EMA updates, both branches adapt directly to similar training signals, reducing the diversity and stability of the predictions used to construct the mixed pseudo-labels. The mean-teacher framework provides a more stable teacher that evolves smoothly over training, complementing the faster-adapting student. This produces more reliable consistency targets and improves segmentation under limited supervision.

\subsection{Effectiveness of Losses}

The loss ablation demonstrates the contribution of each supervision component across both datasets. Using only the partial cross-entropy loss, $L_{pCE}$, results in Dice scores of $0.613$ on CAMUS and $0.410$ on CardiacNet-PAH, showing that sparse scribble annotations contain insufficient supervision information. Adding the consistency loss $L_{cons}$ substantially improves segmentation performance to $0.770$ and $0.626$ respectively, demonstrating the importance of exploiting pseudo-label supervision beyond the annotated pixels. Introducing the temporal consistency loss, $L_{temp}$ further increases Dice to $0.812$ on CAMUS and $0.790$ on CardiacNet-PAH, indicating that coherence between neighboring frames provides valuable supervision. Finally, adding the gated-CRF loss, $L_{CRF}$ improves CAMUS from $0.812$ to $0.851$ but reduces CardiacNet-PAH performance from $0.790$ to $0.781$. This suggests that boundary-aware regularization is beneficial only when image boundaries provide reliable cues.
\section{Conclusion and Future Work}

We propose a weakly-supervised framework for temporal echocardiography segmentation using only two scribbles on an anchor frame. By combining a mean-teacher framework, a LoRA-adapted DINOv3 encoder and connectivity-filtered pseudo-labels with spatial and temporal losses, supervision is propagated across unlabeled sequences. Experiments on CAMUS and CardiacNet demonstrate consistent improvements over existing scribble-supervised methods in Dice and HD95 while substantially reducing false-positive predictions. These results show that annotation cost can be decoupled from sequence length without sacrificing reliable temporal segmentation.
{
    \small
    \bibliographystyle{ieeenat_fullname}
    \bibliography{main}
}

\end{document}